**Properties and device performance of BN thin films grown on GaN by pulsed laser deposition**

Abhijit Biswas,[1,a), b)] Mingfei Xu,[2, b)] Kai Fu,[2,a)] Jingan Zhou,[2] Rui Xu,[1] Anand B. Puthirath,[1] Jordan A. Hachtel,[3] Chenxi Li,[1] Sathvik Ajay Iyengar,[1] Harikishan Kannan,[1] Xiang Zhang,[1] Tia Gray,[1] Robert Vajtai,[1] A. Glen Birdwell,[4] Mahesh R. Neupane,[4] Dmitry A. Ruzmetov,[4] Pankaj B. Shah,[4] Tony Ivanov,[4] Hanyu Zhu,[1] Yuji Zhao,[2,a)] and Pulickel M. Ajayan [1,a)]

**AFFILIATIONS**

[1]Department of Materials Science and Nanoengineering, Rice University, Houston, TX, 77005, USA

[2]Department of Electrical and Computer Engineering, Rice University, Houston, TX, 77005, USA

[3]Center for Nanophase Materials Sciences, Oak Ridge National Laboratory, Oak Ridge, TN 37831, USA

[4] DEVCOM Army Research Laboratory, RF Devices and Circuits, Adelphi, Maryland 20783, USA

[a)]Authors to whom correspondence should be addressed: **ab149@rice.edu, Kai.Fu@rice.edu, Yuji.Zhao@rice.edu, ajayan@rice.edu**

[b)] These authors contributed equally to this work

**ABSTRACT**

Wide and ultrawide-bandgap semiconductors lie at the heart of next-generation high-power, high-frequency electronics. Here, we report the growth of ultrawide-bandgap boron nitride (BN) thin films on wide-bandgap gallium nitride (GaN) by pulsed laser deposition. Comprehensive spectroscopic (core level and valence band XPS, FTIR, Raman) and microscopic (AFM and STEM) characterizations confirm the growth of BN thin films on GaN. Optically, we observed that BN/GaN heterostructure is second-harmonic generation active. Moreover, we fabricated the BN/GaN heterostructure-based Schottky diode that demonstrates rectifying characteristics, lower turn-on voltage, and an improved breakdown capability (~234 V) as compared to GaN (~168 V), owing to the higher breakdown electrical field of BN. Our approach is an early step towards bridging the gap between wide and ultrawide-bandgap materials for potential optoelectronics as well as next-generation high-power electronics.

Wide and ultrawide-bandgap (WBG and UWBG) semiconductors including GaN, SiC, Al(Ga)N, diamond, $Ga_2O_3$, and boron nitride (BN) are expected to revolutionize next-generation electronic device platforms because of their capabilities of realizing lower power loss, smaller system volume, higher operating voltage and temperature, and superior radiation hardness as compared to conventional semiconductors such as Si and GaAs.[1-5] Moreover, heterojunctions between these materials are of great significance for electronic devices as they can modify, and/or further improve the device performance by utilizing the unique advantages of the high break down electric fields, increased thermal conductivity, enhanced carrier mobility, and reduction in interface trap state density, thereby providing improvements in reliability at extreme conditions.[1,2]

Among the aforementioned class of materials, BN is a unique UWBG semiconductor having various structural polymorphs, thus showing diverse properties and potential applications.[6] Specifically, BN is promising for high-power and high-temperature electronics due to its relatively large dielectric constant, UWBG, high breakdown electric fields, high chemical stability, and excellent thermal conductivity.[1] Its most stable polymorph comprises hexagonally stacked layers (h-BN) with a bandgap of ~5.9 eV.[6] The growth of high-quality large-area single-crystalline epitaxial BN thin films remains challenging due to the formation of a vertically disordered h-BN structure, known as turbostratic BN (t-BN).[7-9] There have been several attempts to grow crystalline BN thin films, especially by using metal-organic chemical vapor deposition (MOCVD) on various substrates (e.g. $Al_2O_3$, Si, and metals) showing useful properties and potential applications.[10-13] However, to date, only a few reports exist for BN film growth on III-nitrides (e.g. BN/AlGaN/GaN heterostructures [where BN films were grown via magnetron sputtering]) demonstrating device performances showing low reverse leakage current.[14-17] As BN and GaN are both III-nitrides, the heterojunctions made of BN/GaN can modify and/or improve the GaN-based device performances and could play an important role in high-power electronics and radio-frequency (RF) applications, up to sub-THz regimes.[5,18] Therefore, UWBG/WBG heterostructure-based devices consisting of BN and GaN may open up possibilities for unique high-performance electronics. Besides, owing to its UWBG nature, BN has a large breakdown electrical field that can improve the breakdown performance of GaN Schottky diode devices. Therefore, in this letter, we investigate the properties and device performance of the BN/GaN heterostructures by growing BN thin films on GaN, using pulsed laser deposition (PLD) technique.

Hitherto, PLD has been used to grow BN films (on $c$-Al$_2$O$_3$ and Si) that formed poly, nano-crystalline, and amorphous films.[19-24] For thin film growth, PLD offers some distinct advantages over the MOCVD or sputtering methods.[25] Specifically, the accurate stoichiometry transfer and uniform growth of films from a dense poly- or single-crystalline target leading to epitaxy (depending on the substrate-film lattice mismatch) is due to enhanced adsorbate surface mobility endowed by high energy radicals (both ionized and neutral; B$^+$, N$^+$, N*, N$_2$, N$_2^+$, and BN) that are accelerated from the target in the form of laser-ablated plasma. In addition, relatively low-temperatures (w.r.t. MOCVD) are required for the epitaxial film growth.[25] Considering all these advantages, here we attempt to grow UWBG BN thin films directly on WBG GaN by PLD and demonstrate its various properties and Schottky diode device performance.

BN thin films (~7 nm) were grown by PLD (operating with a UV-KrF excimer laser having 248 nm wavelength and 25 nS pulse width). The films were grown by using the following deposition conditions: growth temperature ~800 °C, laser fluency ~2.2 J/cm$^2$, target-to-substrate distance ~50 mm, repetition rate 5 Hz, and using high-purity nitrogen gas partial pressure (P$_{N2}$) ~100 mTorr. We used a commercially available h-BN target (American Element, 99.9%) for the ablation. For depositions, we used unintentionally doped (0001) GaN (UID-GaN) as a substrate which was grown on hexagonal sapphire $c$-Al$_2$O$_3$ (0001) substrate by MOCVD.[26] Before the growth, GaN substrates were pre-annealed at the same growth temperature (~800 °C) for ~30 min to remove the surface contaminants. For the deposition, 500 laser shots were supplied (thickness of ~7 nm). After the growth, films were cooled down at ~20 °C/min.

To confirm the presence of B and N, we performed X-ray photoelectron spectroscopy (XPS). X-ray photoelectron spectroscopy (XPS) was done using a PHI Quantera SXM scanning X-ray microprobe with a monochromatic Al Kα X-ray source (~1486.6 eV). The C 1s corrected XPS spectra show the presence of dominant B-N peaks, in both the B 1s (~190.7 eV) and N 1s (~398.1 eV) core-level elemental scans [**Figs. 1(a)** and **1(b)**], which are typical characteristics of BN. We also observed small residues of B-O (~192.2 eV), and N-C (~399.5 eV), possibly due to an environmental oxidation effect upon exposure to ambient air and minor-carbonaceous surface contamination.[19,27,28] Interestingly, we also observed the B-B (~188 eV) peak in the XPS, which might be attributed with the replacing B-atoms in the N-site or the formation of boron clusters or via bonding of boron in a hexagonal site to boron in an interstitial site in defecive h-BN.[29,30]

Importantly, the elemental ratio of B:N was found to be close to ~1:1 (33.9:36.0 to be precise), showing the stoichiometric BN growth on GaN.

We characterized the films by Fourier-transform infrared spectroscopy (FTIR) spectroscopy. Fourier-transform infrared spectroscopy (FTIR) spectroscopy was measured by using the Nicolet 380 FTIR spectrometer equipped with a single-crystal diamond window. FTIR spectroscopy of BN provides crucial insight into the morphology and bonding nature. For stacked h-BN sheets, in-plane transverse optical (TO) phonon modes resonate near ~1360 cm$^{-1}$.[31] As shown in FTIR absorption spectra [**Fig. 1(c)**], a strong but broad peak appears in-between 1300-1600 cm$^{-1}$ that is due to the TO vibrations for in-plane B-N bond stretching in $sp^2$ bonded h-BN (ideally at ~1360 cm$^{-1}$). The slight shift in the h-BN-related peak centered towards higher wavenumbers is possibly caused by the defects induced by the high lattice mismatch.[32] We also observed a small peak at ~1200 cm$^{-1}$, attributed to the B-O (~1195 cm$^{-1}$),[32] indicating the presence of adsorbate-related defects on the film surface, in agreement with observation in XPS.

Furthermore, we also obtained Raman spectra by using a Renishaw inVia confocal microscope, operating with a 532 nm laser excitation source that confirms the presence of h-BN related $E_{2g}$ Raman mode with peak center at ~1366 cm$^{-1}$ [**Fig. 1(d)**], with a wide full-width at half-maxima (FWHM) of ~246 cm$^{-1}$. We also observed another peak at ~1600 cm$^{-1}$ that we attributed to the longitudinal optical (LO) $E_{1u}$ infrared active mode caused by the disorder-induced breaking of surface symmetry (related to grain boundaries, edges, disorder, or vacancies) in the BN film.[33-35] Fundamentally, the $E_{2g}$ peak is related to the $sp^2$-bonded modes (long-range B-N ordered) in h-BN.[6] However, our Raman sepctra matches quite well with the reported literature for disordered h-BN.[35] Thus, although the BN film is most likely the disordered/amorphous nature (without any long-range atomic ordering, as shown later by HRTEM), however it may not be "completely" disordered/amorphous. In that case, $sp^2$-bonded regions may exist within the film that gives rise to the observed $E_{2g}$ mode, along with the $E_{1u}$ one.[35] Moreover, we obtained the XPS valence band spectra (VBS) [(**Fig. 1(e)**] exhibiting distinct maxima at ~12 eV (B-N $\pi+\sigma$-band) and ~20 eV (B-N $s$-band) with an additional peak around ~26 eV (O-2$s$), which is consistent with literature reports.[36,37] Furthermore, GaN VBS [inset of **Fig. 1(e)**] also confirms the unintentional $n$-type doping nature.[38] The atomic force microscopy (AFM) surface topography obtained by Park NX20

AFM operating in tapping mode shows the flat surface of GaN and triangular-hape-like morphology of BN film [**Fig. 1(f)**], and exhibits surface roughness of ~0.183 nm and ~0.267 nm, respectively.

To confirm the crystallinity of the constituent layers and resultant heterostructure, we performed cross-sectional atomically resolved high-resolution scanning transmission electron microscopy (STEM) of the film. The TEM specimens were prepared via a focused ion beam (FIB) milling process employing a Helios NanoLab 660 FIB unit. HAADF-STEM images were recorded on a Nion UltraSTEM 100 operating at 100 kV, with a convergence angle of ~32 mrad. The HAADF detector possessed an inner collection angle of ~80 mrad and an outer collection angle of ~200 mrad. The unit was also equipped with a Gatan Enfina electron energy-loss (EEL) spectrometer and the EELS experiments were performed to identify the elemental composition and chemical homogeneity of the sample with an EELS collection semi-angle of ~48 mrad. As can be seen, in-house grown GaN by MOCVD shows single-crystalline nature with atomic ordering [**Fig. 2(a)**]. For the grown BN film, we can clearly see the formation of a ~7 nm dense layer (between Au and GaN); however, it is without any long-range atomic ordering [**Fig. 2(b)**]. The clear presence of BN in the dense layer can be seen with electron-energy-loss spectroscopy (EELS) [**Fig. 2(c)**]. It also shows a reference of the GaN vs. the dense layer, which are then masked and have the total spectra from these regions integrated to show the representative spectra in the EEL spectra [**Fig. 2(d)**]. The K-edges of both B and N differentiate the GaN and BN regions. We can clearly see there is no presence of B in the GaN layer, but it is dominant in the dense layer. In addition, the relative concentrations of B and N are quantified from the EELS mapping of the masked BN layer and found to be in ~1:1 ratio (B:N - 48.3:51.7) [**Figs. 2(e)** and **2(f)**].

Regarding the epitaxy, structurally, to form a perfect layered h-BN stacking, the best epitaxy relationship would be ~4:3 commensurate lattice matching ($4a_{\text{h-BN (0001)}} \approx 3a_{\text{GaN (0001)}}$), between the h-BN ($a = 2.50$ Å, $c = 6.66$ Å) and GaN ($a = 3.18$ Å, $c = 5.18$ Å). However, to maintain this epitaxy relationship, a high ~4.6% compressive strain must be imposed onto the BN. This amount of strain and thus high interfacial lattice energy makes the epitaxial growth energetically unfavorable, at least for the growth conditions used. Hence, it possibly favors the formation of a disordered BN structure to release the interfacial energy, and prefers to grow in Volmer-Weber (three-dimensional island/grain growth) mode without any long-range atomic ordering, as evident through STEM.

The growth of disordered, amorphous-like BN film is also important since it was recently found to be applicable for ultraviolet (UV) photo-detection and high-performance electronics.[39,40]

We performed the optical second-harmonic generation (SHG) on BN film. We imaged the SHG of the BN film with a focused pulsed laser with a center wavelength at 800 nm. The film surface was scanned and the SHG signal with the wavelength around 400 nm was filtered and collected by a single-photon counting module. In general, SHG is important because it allows quantifying the quadratic nonlinear-optical susceptibility ($\chi^{(2)}$) and one of the biggest advantages of BN is its transparent nature throughout the whole visible wavelength spectrum range, potentially opening the doors for visible-range nonlinear optoelectronics. It has been shown that for single-crystalline h-BN, the SHG signal is very strong for the odd number of layers (belongs to the non-centrosymmetric $D_{3h}$ space group), while it goes to zero for an even number of layers.[41,42] In polycrystalline films, the surface roughness (due to random thickness variation), and crystalline orientation at the sub-wavelength scale may result in a statistically finite SHG throughout the sample. Moreover, point and stacking defects in h-BN were reported to enhance the SHG.[43]

**Figure 3(a)** shows a schematic for the reflective SHG excitation and collection from the BN surface. Fundamentally, both (0001) GaN and thick crystalline h-BN are not SHG active for any polarizations according to the crystalline symmetry, although weak remnant SHG was observable from bulk GaN and h-BN, likely due to the imperfect crystalline orientation.[42-44] However, interestingly, our BN/GaN heterostructure showed a uniform SHG signal [**Fig. 3(b)**] that is ~2.2 times of the remnant SHG of GaN. The probability vs. photon intensity count was normalized and shown as a histogram [**Fig. 3(c)**]. From the plot, we obtained the weighted arithmetic mean of BN peak and then calculated the normalized second-order nonlinear optical susceptibility ($\chi^{(2)}$). The calculated net $\chi^{(2)}$ from the BN film was ~0.12 pm/V, which is in close agreement with the previously reported value for the thick ploycrystalline-like h-BN flake.[42] For polycrystalline BN, the effects of domain statistics will give an effect $\chi^{(2)} = \chi^{(2)}_{crys}/sqrt(N)$, where $\chi^{(2)}_{crys}$ is the 2nd nonlinear susceptibility of single crystalline monolayer BN per volume and N is the number of odd-layer BN domains within the SHG emission area. In comparison with the reported $\chi^{(2)}_{crys}$, our measured $\chi^{(2)}$ is two order of magnitudes smaller.[41] This gives us an estimate that there are ~$10^4$ domains inside the SHG emission area (~3 μm in radius), indicating that the domain size is less than 100 nm, which is also seen in the BN AFM image, shown earlier. We also observed stronger

SHG at the sample's edge owning to the symmetry breaking from GaN side surfaces and/or BN's edges.[44,45] The observation of SHG in BN/GaN heterostructure demonstrates a possible way to engineer the tensor components of nonlinear susceptibility forbidden by bulk crystalline symmetry.

We demonstrated the device performance of BN film by fabricating a BN/GaN based Schottky diode and compared its performance with a reference, i.e., an unintentionally doped GaN based Schottky diode. The structures and dimensions of the devices are shown in **Figs. 4 (a)** and **4(b)**. The two kinds of devices were fabricated via the same conventional photolithography and lift-off processes. Metal stacks of Ti (20 nm)/Al (100 nm)/Ti (30 nm)/Au (50 nm) for Ohmic contacts were deposited by electron-beam evaporation in the Nanofab Cleanroom facility at Rice University, followed by rapid thermal annealing at ~850 °C for 30 s under $N_2$ ambient. Metal stacks of Ni (50 nm)/Au (50 nm) for Schottky contacts were deposited also by electron-beam evaporation. The current density-voltage (*J-V*) characteristics were measured by a Keithley 2410 source meter. **Figures 4(c)** and **4(d)** show the *J-V* curves of the two devices in linear scale and semi-log scale, respectively. It can be seen that the forward current of the BN/GaN device is higher than the GaN device, which suggests that the BN layer does not block the current compared with what an insulating layer does in a conventional MIS diode. Meanwhile, we can see a forward leakage current in the BN/GaN device at the beginning of the curve [(**Fig. 4(d)**]. The leakage current can be equivalent to a parallel resistor (P-R). It is necessary to remove the parallel part before we compare the *J-V* characteristics of the devices. As shown in **Fig. 4(e)**, it is clear that the forward current of the BN/GaN device is still higher than the GaN device. So, the 7 nm thick BN insertion layer does not reduce the forward current but increases it and it is more appropriate to consider the BN/GaN device as a Schottky diode rather than an MIS diode.[46] The extracted Schottky barrier height and ideality factor of the GaN diode is ~1.3 eV and ~1.04, respectively, while they are ~1.1 eV and ~1.41 for the BN/GaN diode. The larger ideality factor of the BN/GaN device is supposed to come from the inhomogeneous Schottky barrier and/or interface states due to the defects in the BN by PLD.[47,48] The turn-on voltages for the GaN and BN/GaN diodes are ~1.4 V and ~1.2 V, respectively, by linear extrapolation [(**Fig. 4(f)**]. The relatively lowered Schottky height of the BN/GaN diode not only leads to the increased forward current as compared with the GaN diode but also causes the increased reverse leakage current [**Fig. 5(a)**].

The breakdown voltages of the GaN and BN/GaN diodes are ~168 V and ~234 V, respectively [**Fig. 5(a)**]. To investigate the leakage processes, the reverse *J-V* curves at different temperatures for the devices are shown [**Figs. 5(b)** and **5(c)**]. According to the leakage processes in the Schottky diode,[46] the 3D variable range hopping (VRH) model [$J \propto \exp(-T_c/T)^{1/4}$] has been applied to the leakage in the GaN and BN/GaN devices [**Fig. 5(d)** and **5(e)**]. The ln (*J*) shows a linear relationship versus $T^{-1/4}$, and the characteristic temperatures $T_c$ are calculated to be $1.5 \times 10^8$ K for the GaN device and $3.4 \times 10^5$ K for the BN/GaN device, respectively. These results suggest that the ~7 nm thick BN layer can largely improve the breakdown voltage. This could be due to the higher breakdown electric field of BN than GaN, when subjected to a strong electric field at the edge of the electrode.[23,49]

In summary, we report the growth of UWBG BN thin films on WBG GaN by PLD. Several structural, chemical, spectroscopic, and microscopic characterizations confirm the growth of BN. Functionally, the grown film is second-harmonic generation active, indicating the polar nature. Importantly, the fabricated BN/GaN heterostructure based Schottky diode shows clear rectifying characteristics, lower turn-on voltage, and improved breakdown capability. Our work makes a valuable contribution in the WBG material's domain by exhibiting interesting application potentials in optoelectronics and in high power electronics, thus paving the way in bridging the gap between WBG and UWBG semiconductors.

**ACKNOWLEDGMENTS**

This work was sponsored partly by the Army Research Office and was accomplished under Cooperative Agreement Number W911NF-19-2-0269. It was partly supported as part of ULTRA, an Energy Frontier Research Center funded by the U.S. Department of Energy, Office of Science, Basic Energy Sciences under Award No. DE-SC0021230, and in part by Rice's Technology Development Fund. The views and conclusions contained in this document are those of the authors and should not be interpreted as representing the official policies, either expressed or implied, of the Army Research Office or the U.S. Government. The U.S. Government is authorized to reproduce and distribute reprints for Government purposes notwithstanding any copyright notation herein. Focused ion beam (FIB) milling is performed at the Electron Microscopy Center (EMC) of Rice University and electron microscopy experiments are conducted as part of a user proposal at the Center for Nanophase Materials Sciences, which is a DOE Office of Science User Facility. R.X. and H. Z. are supported by the U.S. National Science Foundation (NSF) under award number DMR 2005096. We would also like to thank the Thermo Fisher Scientific, Oregon USA for help.

## AUTHOR DECLARATIONS

### Conflict of Interest

The authors have no conflicts to disclose.

### Author Contributions

A. B., K. F., P. M. A, and Y. Z. conceptualized the study. A.B., C. L., S. A. I., H. K., X. Z., and T. G. grew and characterized the films. A. B. P. and J. H. performed the FIB and electron microcopy. J. Z., R.X, and H. Z. carried out the optical measurement. K. F., M. X., and Y. Z. fabricated and tested the device. A. G. B., M. R. N., D. A. R., P. B. S, and T. I. commented on the manuscript. All the authors discussed the results and contributed on the manuscript preparation.

## DATA AVAILABILITY

The data that support the findings of this study are available from the corresponding author upon reasonable request.

# FIGURE CAPTIONS

**FIG. 1.** XPS of BN film at (a) B 1s core and (b) N 1s core. (c), (d) FTIR, and Raman spectroscopy show the formation of BN. (e) XPS valence-band spectrum of BN film. Inset shows the VBS of GaN film. (f) AFM surface topography of grown GaN by MOCVD and BN film on GaN by PLD.

**FIG. 2.** Cross-sectional atomically resolved STEM imaging and EELS of the BN film on GaN. (a) HAADF-STEM image of the GaN substrate. (b) Cross sectional bright field (BF)-TEM image of BN film, with GaN substrate on bottom and Au cap on top. (c-f) EELS analysis of BN film. (c) BF-TEM reference of BN-GaN interface (top), which are then masked as GaN (blue) and BN (red) layers (bottom) and the total EELS from these regions are integrated to show the representative spectra of BN and GaN. (d-f) EELS showing the presence of BN film from the masked region (red) in (c), with the presence of B and N in the layer in ~1:1 ratio (B:N - 48.3:51.7).

**FIG. 3.** (a) Schematic of the second-harmonic generation (SHG) excitation and collection from BN surface. (b) Spatial SHG intensity mapping of BN surface. (c) Probability vs. photon intensity counts showing the SHG signal.

**FIG. 4.** Schematic of (a) the GaN and (b) the BN/GaN Schottky diodes. Comparison of *J-V* curves of the two kinds of devices in (c) linear scale and (d) semi-log scale. (e) Comparison of the two kinds of devices after removing the parallel resistance in the BN/GaN diode. (f) Turn-on voltages of the two kinds of devices by linear extrapolation.

**FIG. 5.** (a) Breakdown curves of both the devices at room temperature. Reverse *J-V* curves at different temperatures for (b) the GaN, and (c) the BN/GaN Schottky diodes. The VRH model for the (d) GaN reverse leakage before the breakdown, and for the (e) BN/GaN reverse leakage before the breakdown.

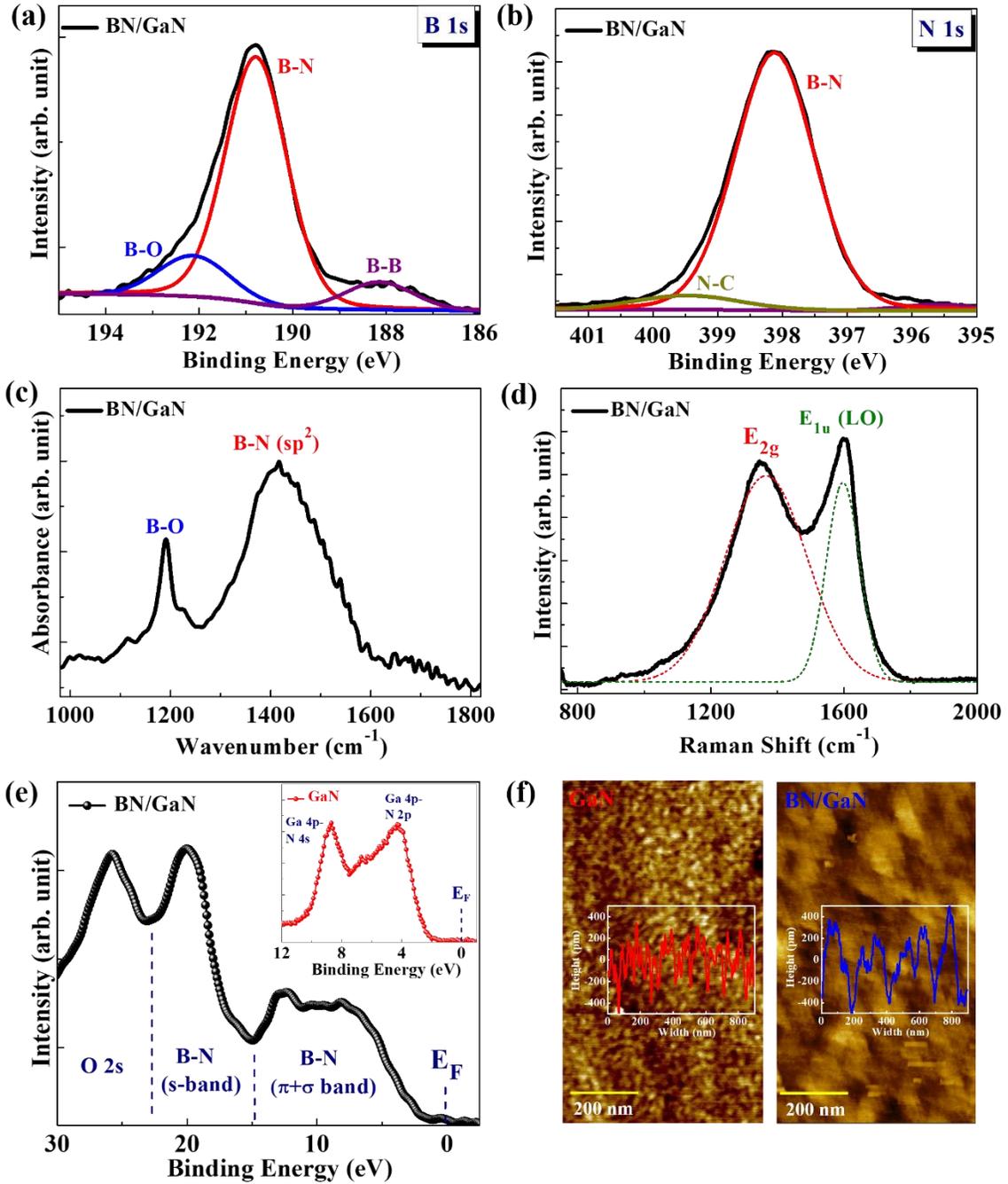

**[FIGURE-2]**

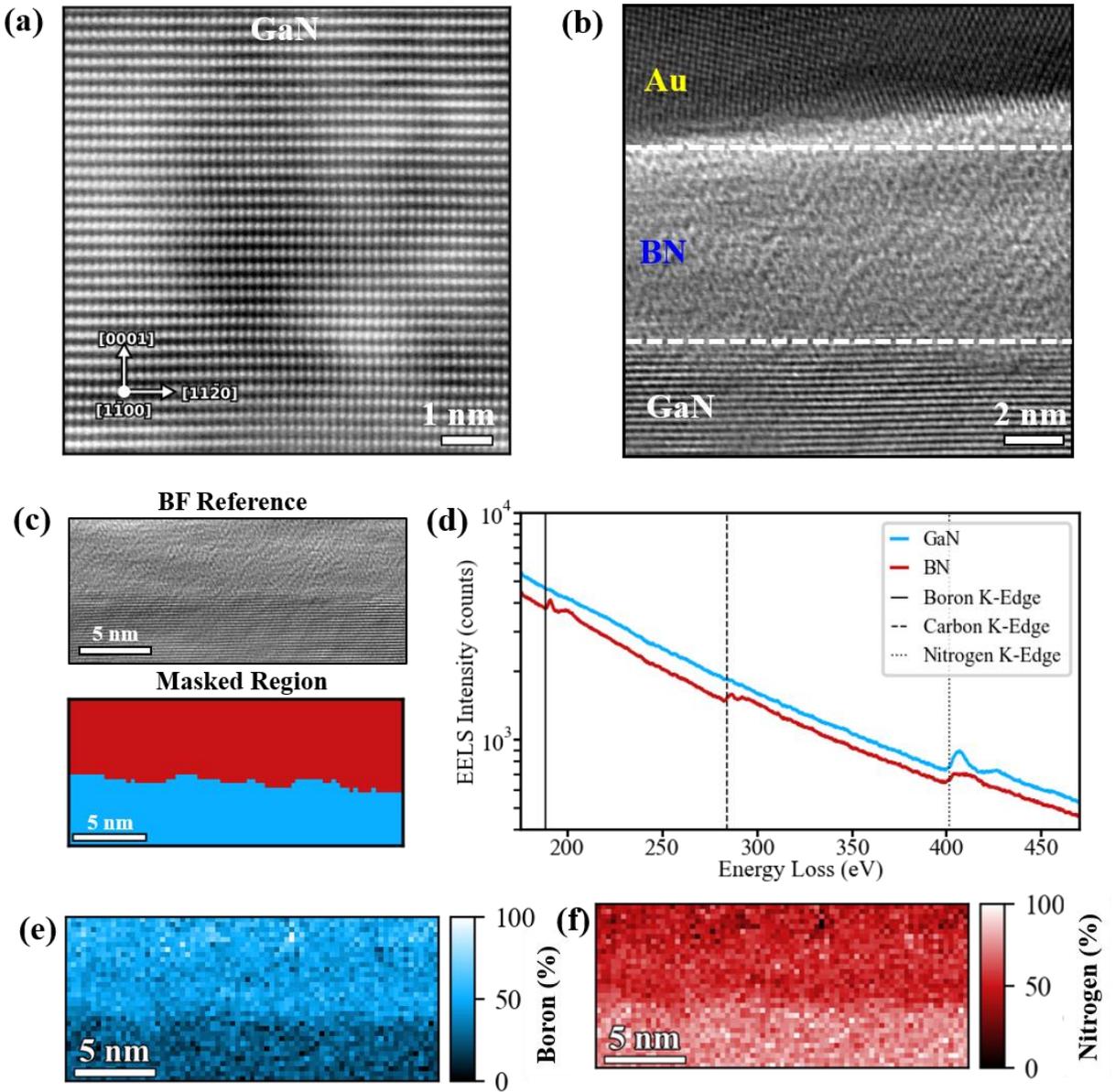

**[FIGURE-3]**

**(a)**

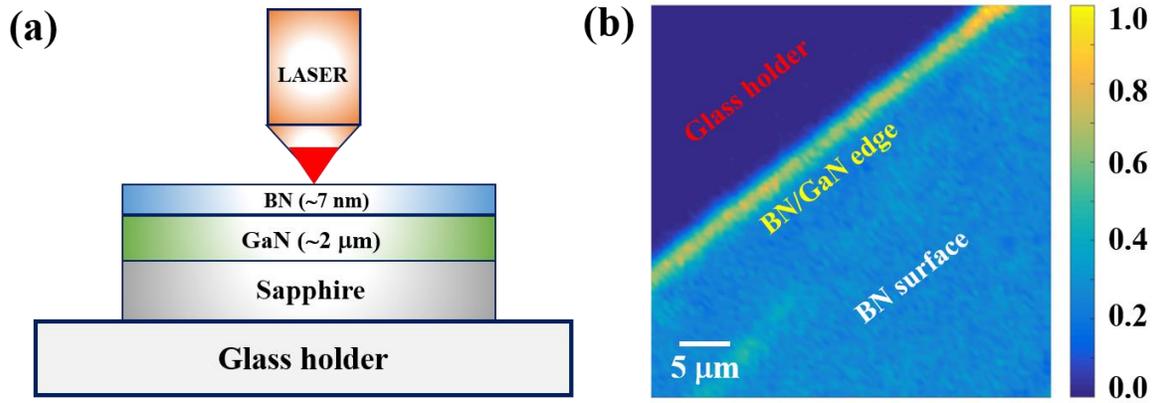

**(c)**

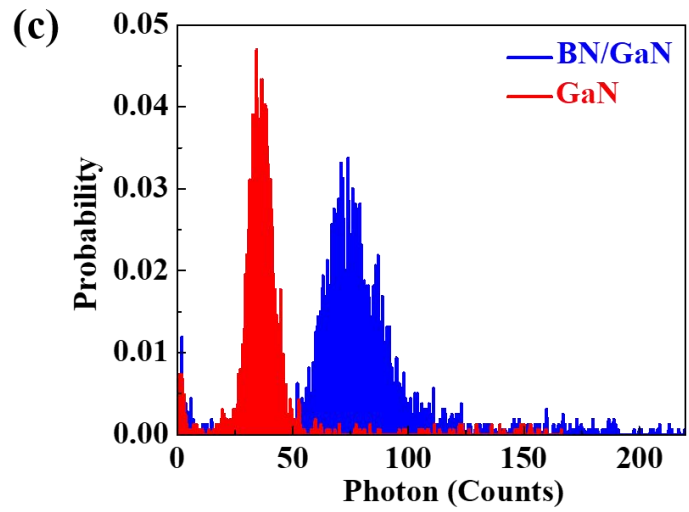

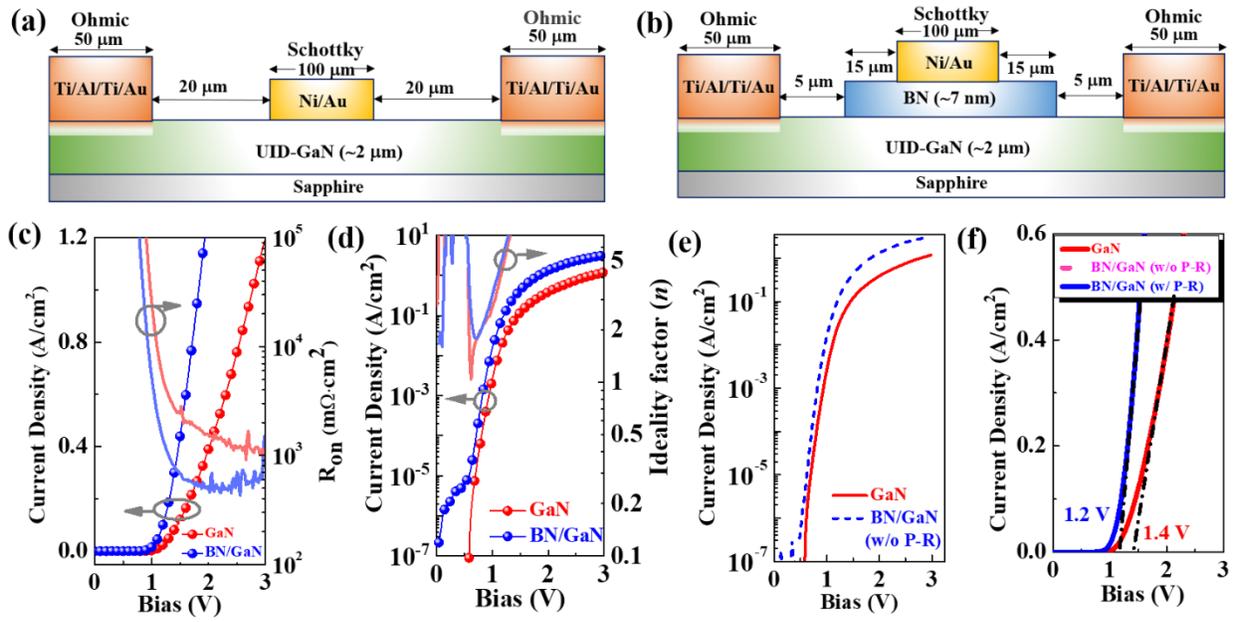

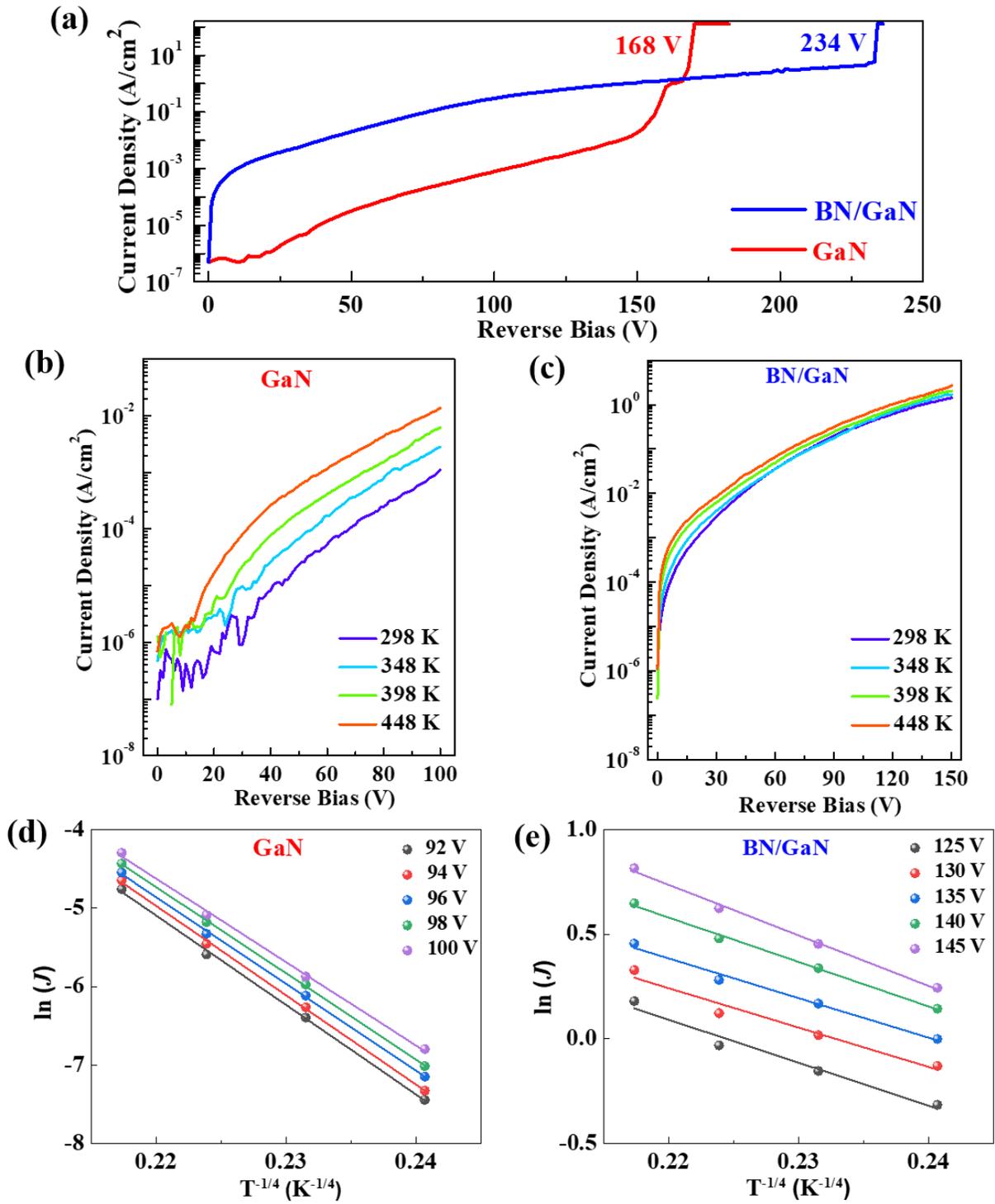